\documentclass{article}

\usepackage{hyperref,url}
\usepackage{amsmath,amssymb,amsthm}

\newtheorem{definition}{Definition}
\newtheorem{proposition}{Proposition}
\newtheorem{lemma}{Lemma}
\newtheorem{remark}{Remark}
\newtheorem{theorem}{Theorem}
\newtheorem{corollary}{Corollary}
\newtheorem{question}{Question}

\title{Grouped Satisficing Paths in Pure Strategy Games: a Topological Perspective}

\author{ 
\begin{tabular}[t]{c c c}
Yanqing Fu &  & Chao Huang \\
Tongji University &  & Tongji University \\
fuyanqing159@163.com & & csehuangchao@tongji.edu.cn \\
& & \\
Chenrun Wang &  & Zhuping Wang \\
Tongji University &  & Tongji University \\
1015239683@qq.com &  & elewzp@tongji.edu.cn \\
\end{tabular}
}

\date{}

\begin{document}

\maketitle

\begin{abstract}
In game theory and multi-agent reinforcement learning (MARL), each agent selects a strategy, interacts with the environment and other agents, and subsequently updates its strategy based on the received payoff. This process generates a sequence of joint strategies $(s^t)_{t \geq 0}$, where $s^t$ represents the strategy profile of all agents at time step $t$. A widely adopted principle in MARL algorithms is "win-stay, lose-shift", which dictates that an agent retains its current strategy if it achieves the best response. This principle exhibits a fixed-point property when the joint strategy has become an equilibrium. The sequence of joint strategies under this principle is referred to as a satisficing path, a concept first introduced in \cite{2023paths} and explored in the context of $N$-player games in \cite{2024paths}. A fundamental question arises regarding this principle: Under what conditions does every initial joint strategy $s$ admit a finite-length satisficing path $(s^t)_{0 \leq t \leq T}$ where $s^0=s$ and $s^T$ is an equilibrium? This paper establishes a sufficient condition for such a property, and demonstrates that any finite-state Markov game, as well as any $N$-player game, guarantees the existence of a finite-length satisficing path from an arbitrary initial strategy to some equilibrium. These results provide a stronger theoretical foundation for the design of MARL algorithms.
\end{abstract}

\section{Introduction}

Game theory provides a formal framework for analyzing strategic interactions among rational decision-makers. It examines how individuals optimize their decisions to maximize their own payoff while accounting for the actions of others. In most settings, agents act in self-interest, leading to the concept of an equilibrium \cite{2019game-theory}, a strategy profile where no agent can unilaterally deviate to achieve a higher payoff. Due to its generality, game theory has become a cornerstone in machine learning, particularly in modeling competitive and cooperative behaviors, such as multi-agent systems \cite{2021multi-agent}, multi-objective reinforcement learning \cite{2013multi-objective}, and adversarial learning \cite{2014gan}.

Multi-agent reinforcement learning (MARL) extends traditional reinforcement learning frameworks to the situation where multiple autonomous agents interact and make decisions concurrently \cite{2023reinforcement-learning}. In multi-agent systems, each agent learns optimal behavior through an iterative process, that is interacting with both the environment and other agents, receiving the reward based on its actions, and dynamically adapting its policy to maximize long-term returns. This paradigm captures the complex inter-dependencies that emerge when multiple learning agents co-evolve within a shared environment.

From the perspective of game theory, MARL can be modeled as a repeated game \cite{2005evolutionary}, where agents iteratively select strategies based on current information, receive rewards from the environment, and update their strategies accordingly. This process generates a strategy path $(s^t)_{t=0}^T$, where $s^t$ represents the joint strategy profile at time step $t$. The concept of equilibrium is particularly crucial in MARL \cite{2001no-regret, 2010on-learning, 2020no-regreet}, as it represents a stable case where no agent can improve its payoff through unilateral deviation. 

A fundamental question in MARL algorithm design is whether decentralized strategy updates can lead the joint strategy to converge to an equilibrium. While this problem has been extensively studied \cite{2000equilibrium-in-sum, 2021adaptive-learning, 2023two-scale}, it remains incompletely resolved \cite{2021marl}. Formally, each agent $i$ updates its strategy according to a revision function: $s_i^{t+1} = f_i(s^u)_{0 \leq u \leq t}$, where $f_i$ maps the history of joint strategies $(s^u)_{0 \leq u \leq t}$ to a new strategy for agent  $i$ \cite{1993statistical, 2012revisiting}. A widely adopted principle in MARL algorithms is "win-stay, lose-shift" \cite{2013aspiration, 2011achieving, 2007global, 2009payoff}: If the agent's current strategy is a best response to other agents, it will maintain this strategy, otherwise it will switch to a substitute strategy. The path generated under this principle is termed a satisficing path \cite{2023paths}.

This paper studies the satisficing path from a topological perspective, aiming to provide new insights into the following core question:

\begin{question} \label{question}
Under what conditions does the game possess the property that for any initial joint strategy $s$, there exists a finite-length satisficing path $(s^t)_{0 \leq t \leq T}$ where $s^0=s$ and $s^T$ is an equilibrium?
\end{question}

This paper adopts a general theoretical perspective to study the existence conditions of the satisficing path, rather than analyzing specific revision methods or update functions $f_i$. The primary objective is to establish fundamental theory of the satisficing path, thereby providing essential theoretical supplements to the existing research on convergence problems in MARL.

\textbf{Contributions.} $1)$ The novel concept of the grouped satisficing path is introduced and formalized in this paper, extending the existing concept of the satisficing path. $2)$ The structural property of the local minimum in the grouped satisficing path is studied from a topological perspective. $3)$ The sufficient conditions for the existence of the grouped satisficing path are established, thus solving Question~\ref{question}. In brief, if the pure strategy game satisfies that $a)$ the strategy set is convex and compact, $b)$ the payoff function is continuous and partially analytic, $c)$ and any sub-game has an equilibrium, then for any initial joint strategy $s$, there exists a finite-length grouped satisficing path $(s^t)_{0 \leq t \leq T}$ where $s^0=s$ and $s^T$ is an equilibrium. More details can be found in Theorem~\ref{theorem:existence}.

This paper is organized as follows. In Section~\ref{section:formalization}, the theoretical framework is established by formalizing key concepts, and the novel concept of the grouped satisficing path is introduced. In Section~\ref{section:topologial-properties}, the topological properties of the grouped satisficing path are studied, and the conditions for reducing the original game to a sub-game are proposed in Theorem~\ref{theorem:topology}. In Section~\ref{section:existence-of-paths}, the sufficient conditions for the existence of the grouped satisficing path are established in Theorem~\ref{theorem:existence}, and this theorem is applied in various situations, such as continuous games, $N$-player games and finite-state Markov games. In Section~\ref{section:discussion}, the theoretical implications of these results are examined, some open problems are discussed for future study, and finally there comes to the conclusion of this paper.

\textbf{Related work.} MARL literature contains numerous algorithms employing the iterative strategy adjustment mechanism. Among these, the fictitious play algorithm \cite{1951iterative} stands as a paradigmatic example that has profoundly influenced development in this field. While extensive research has investigated the convergence properties of the fictitious play and its variants \cite{2006generalised, 2022global-convergence}, existing theoretical guarantees remain limited to the specific game structures \cite{2021best-response, 2022best-response, 2022fictitious-play}. The fundamental connection between the fictitious play and the best-response dynamics deserves particular attention. Agents select optimal strategies based on historical joint strategy profiles, and iteratively update their strategies according to the payoff they get. This formulation suggests that the analytical tools from dynamical systems \cite{2005stochastic, 2005individual-q, 2018best-Response} may offer valuable insights for analyzing the convergence of the game and establishing broader convergence guarantees.

The dynamics of the agent strategies in discrete time settings can be formally characterized by the update equation $s_i^{t+1} = f_i(s^t)$, where each agent $i$'s strategy selection at time step $t+1$ is determined by an update rule $f_i$ that depends on the previous joint strategy profile $s^t$. The seminal work in \cite{2003uncoupled} established a crucial theoretical limitation: for continuous strategy dynamics, if any update function satisfies the regularity conditions and the uncoupled-ness property, the game will be not convergent to an equilibrium generically. Subsequent research \cite{2012completely, 2023impossible} has further generalized these impossibility results. However, notable positive convergence results emerge when introducing stochastic elements. For instance, the agent has the probability to update the strategy randomly when it deviates from the best response \cite{2006stochastic-uncoupled, 2006regret-testing}. The conditional update mechanism has the similar results as the stochastic one \cite{2011convergence-to, 2013near, 2013aspiration}.

The concept of the satisficing path, originally introduced in \cite{2023paths}, enforces a key behavior constraint: agents achieving best response must maintain their current strategies, while allowing unconstrained adaptation for other agents. This flexible formulation naturally encompasses a broad spectrum of learning algorithms \cite{2017decentralized-q, 2018distributed-inertial}. Prior research has progressively established existence guarantees for the satisficing path: \cite{2007global} proved the path existence in two-player games; \cite{2023paths} extended these results to $N$-player symmetric Markov games; \cite{2024paths} demonstrated universal existence in all $N$-player games. While these works successfully addressed the existence problem in specific game classes, they could not extend the result in a more general form directly. So it is necessary to establish the theory of the satisficing path. Since this paper studies the topological properties of the satisficing path, the existence results can be successfully extended in a more general form, thus solving Question~\ref{question} mentioned above.

\section{Formalization} \label{section:formalization}

\subsection{Pure strategy game}

A pure strategy game is defined by the tuple $G=(I, (S_i)_{i\in I}, (g_i)_{i \in I})$, where $I$ represents the set of players, $S_i$ denotes the pure strategy set of player $i$, and $g_i: \prod_{i \in I} S_i \to \mathbb{R}$ is player $i$'s payoff function. In this paper, only pure strategy games with finite players ($\vert I \vert < \infty$) are discussed.

Each player $i \in I$ independently chooses a strategy $s_i \in S_i$ from their respective strategy set. The resulting payoff for player $i$ is determined by the joint strategy profile $(s_1, s_2, \ldots, s_{\vert I \vert})$ through the payoff function $g_i$. Each rational player aims to maximize their individual payoff. However, since the payoff functions depend on all players' simultaneous strategy choices and the players cannot directly control others' strategy selections, the optimal strategic choice reduces to selecting a best response to the current strategies of other players.

\begin{definition} \label{definition:best-response}
In a pure strategy game $G$, the $\epsilon$-best response correspondence for player $i$ given opponents' strategy profile $s_{-i}$ is
\begin{equation}
BR_{\epsilon}(s_{-i}) = \left\{s \in S_i \mid g_i(s, s_{-i}) \geq \sup_{t \in S_i} g_i(t, s_{-i}) - \epsilon \right\}.
\end{equation}
where $s_{-i} \in \prod_{j \neq i, j \in I} S_j$ denotes the joint strategy of all other players. In particular, the notation $BR_{\epsilon}$ will be simplified to $BR$ when $\epsilon=0$.
\end{definition}

In the absence of additional constraints, the best response set $BR(s_{-i})$ may be empty, as the existence of a maximum point in $S_i$ is not guaranteed a priori. When $S_i$ is closed and compact, and $g_i$ is continuous, then $BR(s_{-i})$ is non-empty for all $s_{-i} \in S_{-i}$. For $\epsilon > 0$, the existence condition can be relaxed to the boundedness of $g_i$.

\begin{definition} \label{definition:equilibrium}
In a pure strategy game $G$, a joint strategy $s \in \prod_{i \in I} S_i$ is called an $\epsilon$-equilibrium if
\begin{equation}
\forall i \in I, \quad s_{i} \in BR_{\epsilon}(s_{-i}).
\end{equation}
\end{definition}

A mixed strategy game can also be defined by the tuple $G=(I, (S_i)_{i\in I}, (g_i)_{i \in I})$. However, each player $i\in I$ can choose a Borel probability measure over $S_i$ as a mixed strategy. Construct a pure strategy game $\tilde{G}=(I, (\Delta S_i)_{i\in I}, (\tilde{g}_i)_{i \in I})$, where $\Delta S_i$ denotes the set of all Borel probability measures over $S_i$, and
$$
\tilde{g}_i: \prod_{i \in I} \Delta S_i \to \mathbb{R}, \quad \prod_{i \in I} \sigma_i \mapsto \int_{s \in \prod_{i \in I} S_i} g_i(s) \prod_{i\in I} d\sigma_i(s_i).
$$
Then the equilibrium of the pure strategy game $\tilde{G}$ is equivalent to the Nash equilibrium of the original mixed strategy game $G$. This fundamental equivalence justifies referring to $\tilde{G}$ as the extended game of $G$.

\begin{remark}
The class of pure strategy games exhibits strict representational generality over mixed strategy games. For any mixed strategy game, there exists an equivalent pure strategy game which possesses the same properties of the original game. However, a pure strategy game admits a mixed strategy representation only if each strategy set can be embedded into the set of Borel probability measures over some base space. 
\end{remark}

\subsection{Grouped satisficing path}

In a pure strategy game $G$, the repeated play generates a path $(s^t)_{t \geq 0}$ where $s^t$ represents the joint strategy profile at time step $t$. In this paper, only discrete time strategy dynamics ($t \in \mathbb{N}$) is discussed.

\begin{definition} \label{definition:satisficing-path}
In a pure strategy game $G$, the strategy path $(s^t)_{t \geq 0}$ is called an $\epsilon$-satisficing path if
\begin{equation}
\forall t \geq 0, \quad \forall i \in I, \quad s_i^t \in BR_{\epsilon}(s_{-i}^t) \quad \Rightarrow \quad s_i^{t+1} = s_i^t.
\end{equation}
\end{definition}

The definition of the satisficing path is an accurate description of the "win-stay, lose-shift" principle. If the player's strategy is a current best response, it will maintain this strategy in the next time step, otherwise it will explore an alternative strategy to replace it. This principle is fundamental in the MARL algorithm design, since it ensures the stability at the equilibrium.

\begin{definition} \label{definition:grouped-satisficing-path}
In a pure strategy game $G$, the group set $P$ is a partition of the player set $I$ satisfying:
\begin{enumerate}
\item $\bigcup_{p \in P} p = I$.
\item For any $p,q \in P$, if $p \neq q$, then $p \cap q = \varnothing$.
\end{enumerate}
A strategy path $(s^t)_{t \geq 0}$ is called a grouped $\epsilon$-satisficing path with respect to the group set $P$ if
\begin{equation}
\forall t \geq 0, \quad \forall p \in P, \quad (\forall i \in p, \, s_i^t \in BR_{\epsilon}(s_{-i}^t)) \quad \Rightarrow \quad (\forall i \in p, \, s_i^{t+1} = s_i^t).
\end{equation}
\end{definition}

The definition of the grouped satisficing path extends the "win-stay, lose-shift" principle to player groups. The basic decision unit is a group of players rather than individual players. If all players in a group achieve best responses, all of them will maintain their strategies. When each group contains exactly one player ($P = \{\{i\} \mid i \in I\}$), the grouped satisficing path reduces to the classical satisficing path.

\begin{remark}
The grouped satisficing path constitutes a proper generalization of the classical satisficing path. Obviously, any satisficing path can be viewed as a grouped satisficing path, while the converse proposition is not always true. The grouped satisficing path serves as an essential technical tool for proving the existence of the satisficing path in finite-state Markov games.
\end{remark}

\begin{definition} \label{definition:number}
In a pure strategy game $G$ with the group set $P$, for a joint strategy $s$, the $\epsilon$-best response group count with respect to $s$ is
\begin{equation}
N_{\epsilon}(s) = \vert \{p \in P \mid \forall i \in p, \, s_i \in BR_{\epsilon}(s_{-i}) \} \vert.
\end{equation}
In particular, the notation $N_{\epsilon}$ will be simplified to $N$ when $\epsilon=0$.
\end{definition}

The $\epsilon$-best response group count constitutes an important characteristic of joint strategy profiles. It quantifies the number of player groups where all members simultaneously achieve $\epsilon$-best responses. 

\begin{definition} \label{definition:local}
In a pure strategy game $G$ with the group set $P$, for a joint strategy $s$, the admissible subsequent joint strategy set with respect to $s$ is
\begin{equation}
T_{\epsilon}(s) = \left\{ s' \in \prod_{i \in I} S_i \mid (s, s') \text{ satisfies the grouped $\epsilon$-satisficing dynamics} \right\}.
\end{equation}
$N_{\epsilon}(s)$ is called a local minimum if
\begin{equation}
N_{\epsilon}(s) \leq \inf_{t \in T_{\epsilon}(s)} N_{\epsilon}(t).
\end{equation}
$N_{\epsilon}(s)$ is called a local maximum if
\begin{equation}
N_{\epsilon}(s) \geq \sup_{t \in T_{\epsilon}(s)} N_{\epsilon}(t).
\end{equation}
In particular, the notation $T_{\epsilon}$ will be simplified to $T$ when $\epsilon=0$.
\end{definition}

The admissible subsequent joint strategy set $T_{\epsilon}(s)$ is the set of all possible joint strategies following $s$ without violating the property of grouped $\epsilon$-satisficing path. So, a local minimum means that any admissible joint strategy cannot decrease the $\epsilon$-best response group count, while maximum means cannot increase.

\subsection{Two critical definitions}

\begin{definition} \label{definition:analytic}
$\{X_i\}_{i\in I}$ is a family of sets, where each $X_i$ is a convex set in a topological vector space. A function $f: \prod_{i \in I} X_i \to \mathbb{R}$ is called a partially analytic function if for any $i \in I$, $p_i,q_i \in X_i$, $x_{-i} \in \prod_{j \neq i, j \in I} X_j$, the function
$$
g: [0, 1] \to \mathbb{R}, \quad a \mapsto f(ap_i + (1-a)q_i, x_{-i}).
$$
is real analytic with respect to $a$.
\end{definition}

\begin{definition} \label{definition:sub-game}
In a pure strategy game $G$ with the group set $P$, it is called that any sub-game has an equilibrium, if for any group subset $Q \subset P$ and any joint strategy $s$, the sub-game $\tilde{G} = (J, (S_i)_{i \in J}, (\tilde{g}_i)_{i \in J} )$ admits at least one equilibrium, in which $J = \bigcup_{p \in Q} p$ and 
$$
\tilde{g}_i: \prod_{j\in J} S_j \to \mathbb{R}, \quad w \mapsto g_i(s_{-J}, w).
$$
where $s_{-J} = (s_k)_{k \in I \backslash J}$ denotes the strategy profile of all players outside $J$.
\end{definition}

The two definitions above represent the two conditions stated in Theorem~\ref{theorem:existence}.

\section{Topological properties} \label{section:topologial-properties}

Two fundamental lemmas are established first and form the basis for proving the main theorem of the topological properties. Lemma~\ref{lemma:Baire} provides the necessary conditions for a joint strategy $s$ to be a local minimum of the $\epsilon$-best response group count $N_{\epsilon}(s)$. Lemma~\ref{lemma:analytic} characterizes the local behavior of the best response when a neighborhood of the strategy satisfies optimality conditions. Theorem~\ref{theorem:topology} provides the sufficient conditions under which the original game can be reduced to a well-defined sub-game. Due to space limitation, the detailed proofs of Lemma~\ref{lemma:Baire} and Lemma~\ref{lemma:analytic} appear in the Appendix. 

\begin{lemma} \label{lemma:Baire}
In a pure strategy game $G$ with the group set $P$, suppose that each $S_i$ is compact and Hausdorff, and each $g_i$ is continuous. For a joint strategy $s$, if $N_{\epsilon}(s)$ is a local minimum, then

\begin{enumerate}
\item either for any joint strategy $t \in T_{\epsilon}(s)$ in the admissible subsequent joint strategy set, the group which has achieved $\epsilon$-best response in $s$ will still achieve $\epsilon$-best response in $t$, 
\item or there exists a non-empty open set $O \subset T_{\epsilon}(s)$ in the admissible subsequent joint strategy set satisfying:
\begin{enumerate}
\item For any joint strategy $t \in O$, there exists a group which has achieved $\epsilon$-best response in $s$ will not achieve $\epsilon$-best response in $t$.
\item There exists a group which has not achieved $\epsilon$-best response in $s$ will achieve $\epsilon$-best response in any joint strategy $t \in O$.
\end{enumerate}
\end{enumerate}

\end{lemma}

Lemma~\ref{lemma:Baire} is proved in Appendix~\ref{appendix:Baire}. In brief, the main idea is to construct a finite family of closed sets covering a given Baire space, and then to discuss different cases about this Baire space. The property of the Baire space will be used in this proof.

\begin{lemma} \label{lemma:analytic}
In a pure strategy game $G$, suppose that each $S_i$ is a convex set in a topological vector space, and each $g_i$ is partially analytic. If there exists a non-empty open set $U \subset \prod_{i \in I} S_i$ satisfying that the player $k$ achieves best response in any joint strategy $s \in U$, then the player $k$ will achieve best response in any joint strategy $s \in \prod_{i \in I} S_i$.
\end{lemma}

Lemma~\ref{lemma:analytic} is proved in Appendix~\ref{appendix:analytic}. In brief, the main idea is to construct a function to describe the property of the best response, and then to use the analytic extendability to expand the best response set step by step.

\begin{theorem} \label{theorem:topology}
In a pure strategy game $G$ with the group set $P$, suppose that each $S_i$ is a convex compact set in a topological vector space, and each $g_i$ is continuous and partially analytic. Then, for a joint strategy $s$, the best response group count $N(s)$ is a local minimum, if and only if, for any joint strategy $t \in T(s)$ in the admissible subsequent joint strategy set, the group which has achieved best response in $s$ still achieves best response in $t$.
\end{theorem}

\begin{proof}

\textbf{Sufficiency.} Suppose for any joint strategy $t \in T(s)$ in the admissible subsequent joint strategy set, the group which has achieved best response in $s$ still achieves best response in $t$. This means that for any joint strategy $t \in T(s)$, there exists $N(s) \leq N(t)$. As a result
$$
 N(s) \leq \inf_{t \in T(s)} N(t).
$$
According to Definition~\ref{definition:local}, $N(s)$ is a local minimum.

\textbf{Necessity.} Suppose the best response group count $N(s)$ is a local minimum. Since $S_i$ is the set in a topological vector space, $S_i$ is Hausdorff. According to Lemma~\ref{lemma:Baire}, there are two cases:

$1)$ Either any joint strategy in $T(s)$ makes the group which has achieved best response in $s$ will still achieve, 

$2)$ or there exists a non-empty open set $O\subset T(s)$ satisfying: $a)$ There exists one group $p \in P$ which has not achieved best response in $s$ will achieve in any joint strategy in the set $O$, $b)$ and for any joint strategy in the set $O$, there exists one group which has achieved best response in $s$ will not achieve.

Suppose case $2)$ is valid. Assume group $p_1,\ldots,p_n$ do not achieve best response in $s$. Obviously, $p \in  \{p_1, \ldots, p_n\}$. Let $Q=\bigcup_{i=1}^n p_i$. Construct a pure strategy game $\tilde{G}=(Q, (S_i)_{i\in Q}, (\tilde{g}_i)_{i \in Q})$. $\tilde{g}_i$ is defined as
$$
\tilde{g}_i: \prod_{j \in Q} S_j \to \mathbb{R}, \quad w \mapsto g_i(s_{-Q}, w).
$$
where $s_{-Q} = (s_k)_{k \in I \backslash Q}$ denotes the strategy profile of all players outside $Q$.

Let the projection of the non-empty open set $O$ onto $\prod_{j\in Q} S_j$ be $O_p$. Obviously, $O_p$ is a non-empty open set, and each player in $p$ achieves best response in any joint strategy in $O_p$. Each $\tilde{g}_i$ is actually a constrained function of $g_i$. Since $g_i$ is continuous and partially analytic, $\tilde{g}_i$ is also continuous and partially analytic. According to Lemma~\ref{lemma:analytic}, for any joint strategy $w \in \prod_{j \in Q} S_j$, each player in $p$ will achieve best response in $w$. In turn, for any joint strategy $t \in T(s)$, each player in $p$ will achieve best response in $t$ in the game $G$. However, $s \in T(s)$. This is a contradiction with the fact that group $p$ does not achieve best response in $s$. So case $2)$ is invalid.

As a result, for any joint strategy $t \in T(s)$, the group which has achieved best response in $s$ will still achieve best response in $t$.
\end{proof}

\begin{remark}
This theorem actually establishes a formal reduction principle for games under grouped satisficing dynamics. When there exists a group where all players fix their strategies as their strategies are always best responses in all possible grouped satisficing paths, it can be formally treated as part of the environmental dynamics, reducing the original game to a sub-game.
\end{remark}

\section{Existence of paths} \label{section:existence-of-paths}

In this section, a fundamental existence theorem for the grouped satisficing path in pure strategy games is established. Theorem~\ref{theorem:existence} provides the sufficient conditions for the existence, thus answering Question~\ref{question}. This theorem yields four corollaries, each addressing distinct game-theoretic scenarios. Due to space limitation, the proofs of these corollaries appear in the Appendix. 

\begin{proposition} \label{proposition:finite-minimum}
In a pure strategy game $G$ with the group set $P$, for any infinite-length path $(s^t)_{t \geq 0}$, there exists an index $v$ satisfying $N_{\epsilon}(s^v) = \inf_{t \geq 0} N_{\epsilon}(s^t)$. 
\end{proposition}

Proposition~\ref{proposition:finite-minimum} is proved in Appendix~\ref{appendix:existence-of-paths}.

\begin{theorem} \label{theorem:existence}
In a pure strategy game $G$ with the group set $P$, suppose that each $S_i$ is a convex compact set in a topological vector space, each $g_i$ is continuous and partially analytic, and any sub-game has an equilibrium. Then for any initial joint strategy $s$, there exists a finite-length grouped satisficing path $(s^t)_{t=0}^T$ where $s^0=s$ and $s^T$ is an equilibrium.
\end{theorem}

\begin{proof}
Any finite-length grouped satisficing path $(s^t)_{t=0}^T$ where $s^T$ is an equilibrium can be extended to an infinite-length one. Construct a path
$$
a^t = \begin{cases}
s^t, & 0 \leq t \leq T, \\
s^T, & T < t.\\
\end{cases}
$$
Since $a^T$ is an equilibrium, any player will not change their strategies any longer. So $(a^t)_{t \geq 0}$ is a grouped satisficing path.

Construct a set
$$
A = \{(s^t)_{t \geq 0} \mid (s^t)_{t \geq 0} \text{ is an infinite-length grouped satisficing path}, \, s^0=s\}.
$$
According to Proposition~\ref{proposition:finite-minimum}, each path $(s^t)_{t \geq 0}$ has a minimum of $N(s^t)$. Construct $\vert P \vert +1$ sets
$$
A_i = \left\{(s^t)_{t \geq 0} \in A \mid \inf_{t \geq 0} N(s^t) = i\right\}, \quad i=0,\ldots, \vert P \vert.
$$
There exists a minimum $k$ satisfying $A_k \neq \varnothing$. Choose a path $(s^t)_{t \geq 0} \in A_k$. According to Proposition~\ref{proposition:finite-minimum}, there exists $s^x \in (s^t)_{t \geq 0}$ satisfying $N(s^x) = k$.

Reduction to absurdity needs to be used here.

\textbf{Assume on the contrary that} $N(s^x)$ is not local minimum. By Definition~\ref{definition:local}, there exists a joint strategy $u$ following $s^x$ without violating the property of grouped satisficing path, and satisfying $N(u) \leq N(s^x) -1$. Construct a path
$$
a^t = \begin{cases}
s^t, & 0 \leq t \leq x, \\
u, & x < t.\\
\end{cases}
$$
It is sure that $(a^t)_{t=0}^{x+1}$ is a grouped satisficing path. For any $t>x$, any player does not change their strategies, which means that the group who has achieved best response will still achieve, and the group who not will still not. So $(a^t)_{t \geq 0}$ is actually a grouped satisficing path. As a result, $(a^t)_{t \geq 0} \in A$, and
$$
\inf_{t \geq 0} N(a^t) \leq N(u) \leq N(s^x) -1 = k-1.
$$
So $(a^t)_{t \geq 0}$ belongs to one of $A_0, \ldots, A_{k-1}$, which is a contradiction with the fact that $A_k$ is the non-empty set with the minimal index.

As a result, $N(s^x)$ is local minimum.

According to Theorem~\ref{theorem:topology}, any possible subsequent joint strategy following $s^x$ without violating the property of grouped satisficing path, makes the group which has achieved best response in $s^x$ will still achieve. Assume group $p_1,\ldots,p_n$ do not achieve best response in $s^x$. Let $J=\bigcup_{i=1}^n p_i$ Construct a pure strategy game $\tilde{G}=(J, (S_i)_{i\in J}, (\tilde{g}_i)_{i \in J})$. $\tilde{g}_i$ is defined as
$$
\tilde{g}_i: \prod_{j \in J} S_j \to \mathbb{R}, \quad w \mapsto g_i(s^x_{-J}, w).
$$
where $s^x_{-J} = (s^x_k)_{k \in I \backslash J}$ denotes the strategy profile of all players outside $J$.

By assumption that any sub-game has an equilibrium, there exists an equilibrium $w$ in $\tilde{G}$. Let
\begin{equation} \label{equation:define-u}
u = (s^x_{-J}, w) \in \prod_{i\in I} S_i.
\end{equation}
Construct a finite path
$$
a^t = \begin{cases}
s^t, & 0 \leq t \leq x, \\
u, & t=x+1.\\
\end{cases}
$$
Obviously, $\{a^t\}_{t=0}^{x+1}$ is a grouped satisficing path, and $a^0 = s$. According to Theorem~\ref{theorem:topology} and Equation~\ref{equation:define-u}, the group which has achieved best response in $s^x$ will still achieve in $u$. Since $w$ is an equilibrium in the sub-game $\tilde{G}$, the group which has not achieved best response in $s^x$ will achieve in $u$. As a result, $a^{x+1}$ is an equilibrium in $G$.

So $\{a^t\}_{t=0}^{x+1}$ is a finite-length grouped satisficing path where $a^0=s$ and $a^{x+1}$ is an equilibrium.

\end{proof}

This theorem establishes fundamental guarantees for the existence of the grouped satisficing path. While the technical requirements may appear stringent, they are in fact satisfied in numerous practical scenarios.

\begin{corollary} \label{corollary:continuous-game}
In a pure strategy game $G$, suppose that each $S_i$ is a convex compact set in a topological vector space and each $g_i$ is continuous and partially analytic. If for any $g_i$ and any $s_{-i} \in \prod_{j \neq i, j\in I} S_j$, the function $g_i(s_i, s_{-i})$ is quasi-convex with respect to $s_i$, then for any initial joint strategy $s$, there exists a finite-length satisficing path $(s^t)_{t=0}^T$ where $s^0=s$ and $s^T$ is an equilibrium.
\end{corollary}

Corollary~\ref{corollary:continuous-game} is proved in Appendix~\ref{appendix:continuous-game}. The assumption of this corollary is almost the same as Theorem~\ref{theorem:existence}, so only need to check whether any sub-game has an equilibrium. 

\begin{corollary} \label{corollary:simple-game}
In a mixed strategy game $G$, suppose that each $S_i$ is a finite set. Then for any initial joint mixed strategy $\sigma$, there exists a finite-length satisficing path $(\sigma^t)_{t=0}^T$ where $\sigma^0=\sigma$ and $\sigma^T$ is a mixed equilibrium.
\end{corollary}

Corollary~\ref{corollary:simple-game} is proved in Appendix~\ref{appendix:simple-game}. This corollary is essentially the main result of \cite{2024paths}, where the authors directly construct a satisficing path and prove that the equilibrium is a limit point under certain conditions. In contrast, the existence of such a path in $N$-player games is established as a corollary of Theorem~\ref{theorem:existence} in this paper. Thus, the proof reduces to verifying that the conditions of Theorem~\ref{theorem:existence} hold in this setting.

A stationary mixed strategy stochastic game is defined as $G=(I, (S_i)_{i \in I}, X, P, (g_i)_{i\in I}, (\gamma_i)_{i\in I})$. $I$ is the set of players. $S_i$ is the strategy set of player $i$. $X$ is the state set. $P: X \times \prod_{i \in I} S_i \to \Delta X $ is the transition probability function, mapping the current state and joint strategy to a probability distribution over the next states. $g_i: X \times \prod_{i \in I} S_i \to \mathbb{R}$ is the payoff function for player $i$. $\gamma_i \in [0,1)$ is the discount factor for player $i$. In this paper, only stationary mixed strategy stochastic games with finite players and finite states ($\vert I \vert < \infty, \vert X \vert < \infty$) are discussed.

Each player $i$ selects a strategy from $S_i$ according to the probability distribution $\pi_i(x)$. The mapping $\pi_i: X \to \Delta S_i$ is called a stationary mixed strategy, as the player $i$'s strategy depends solely on the current state $x \in X$. Given a joint strategy profile $s$ and the current state $x$, two events occur: $1)$ Each player $i$ receives an immediate payoff $g_i(x,s)$, $2)$ and the game transits to a new state $x' \in X$ with probability $P(x,s)(x')$. As the game progresses, the player $i$ obtains a sequence of payoffs $\{g_i(x^t,s^t)\}_{t \geq 0}$ and aims to maximize the discounted average payoff $\sum_{t \geq 0} \gamma_i^t g_i(x^t,s^t)$.

A joint mixed strategy profile $(\pi_i)_{i \in I}$ constitutes a mixed equilibrium in a stationary mixed strategy stochastic game, if for any initial state $x \in X$ and any player $i \in I$, there exists
$$
\mathbb{E}_{(\pi_j)_{j \in I}} \left[ \sum_{t \geq 0} \gamma_i^t g_i(x^t,s^t) \mid x_0=x \right] = \sup_{\sigma_i \in (\Delta S_i)^X} \mathbb{E}_{\sigma_i, (\pi_j)_{j\neq i,j \in I}} \left[ \sum_{t \geq 0} \gamma_i^t g_i(x^t,s^t) \mid x_0=x \right].
$$

\begin{corollary} \label{corollary:stationary-game}
In a stationary mixed strategy stochastic game $G$, suppose that each $S_i$ is a finite set. Then for any initial stationary joint mixed strategy $\sigma$, there exists a finite-length satisficing path $(\sigma^t)_{t=0}^T$ where $\sigma^0=\sigma$ and $\sigma^T$ is a mixed equilibrium.
\end{corollary}

Corollary~\ref{corollary:stationary-game} is proved in Appendix~\ref{appendix:stationary-game}. The proof of this corollary is a bit technically demanding. First, it is necessary to transform the stationary mixed strategy stochastic game with the satisficing path into a pure strategy game with the grouped satisficing path. Since it contains infinite summation, the validity of the operation must be verified. Then, check whether each condition listed in Theorem~\ref{theorem:existence} is satisfied.  Contraction mapping theorem will be used in this proof.

Similar to the definition of the stationary mixed strategy stochastic game, a $k$-step mixed strategy stochastic game can also be defined as $G=(I, (S_i)_{i \in I}, X, P, (g_i)_{i\in I}, (\gamma_i)_{i\in I})$, where all components are defined analogously to the stationary mixed strategy case. However, each player $i$ selects a strategy from $S_i$ according to the probability distribution $\pi_i(x, s^{-1}, \ldots, s^{-k})$, where $s^{-t}$ denotes the joint strategy profile from $t$ steps before the current time step. So $\pi_i : X \times ( \prod_{i\in I} S_i )^k \to \Delta S_i$. Similarly, only the case with finite players and finite states ($\vert I \vert < \infty, \vert X \vert < \infty$) is discussed in this paper.

\begin{corollary} \label{corollary:k-step-game}
In a $k$-step mixed strategy stochastic game $G$, suppose that each $S_i$ is a finite set. Then for any initial stationary joint mixed strategy $\sigma$, there exists a finite-length satisficing path $(\sigma^t)_{t=0}^T$ where $\sigma^0=\sigma$ and $\sigma^T$ is a mixed equilibrium.
\end{corollary}

Corollary~\ref{corollary:k-step-game} is proved in Appendix~\ref{appendix:k-step-game}. The main idea is to find a bijection between the paths in the $k$-step mixed strategy stochastic game and the paths in a stationary mixed strategy stochastic game, and then to use Corollary~\ref{corollary:stationary-game} to come to the conclusion.

\begin{remark}
Usually, the game in Corollary~\ref{corollary:continuous-game} is called a continuous game, where the payoff function is continuous. The game in Corollary~\ref{corollary:simple-game} is termed an $N$-player game, while the games in Corollary~\ref{corollary:stationary-game} and Corollary~\ref{corollary:k-step-game} are referred to as finite-state Markov games. Corollary~\ref{corollary:simple-game} establishes the existence of the satisficing path in a general $N$-player game, Corollary~\ref{corollary:stationary-game} proves its existence in a standard reinforcement learning setting, and Corollary~\ref{corollary:k-step-game} extends the result to the reinforcement learning with historical records.
\end{remark}

\section{Discussion} \label{section:discussion}

Theorem~\ref{theorem:topology} establishes a framework in which the grouped satisficing paths of the original game can be derived from those of the sub-game. Consequently, the original game can be effectively reduced to its sub-game. Theorem~\ref{theorem:existence} presents sufficient conditions for the existence of the grouped satisficing paths connecting any initial joint strategy to an equilibrium. Thus, Theorem~\ref{theorem:existence} provides a solution to Question~\ref{question}, or more precisely, it identifies sufficient conditions for the question's resolution.

From a practical standpoint, the corollaries of Theorem~\ref{theorem:existence} may be more valuable. Specifically, these results demonstrate that: $1)$ In any $N$-player game, there exists satisficing paths from any initial joint mixed strategy to a mixed equilibrium. $2)$ For any finite-state Markov game, whether stationary or finite-step, such satisficing paths always exist from arbitrary initial joint mixed strategy to a mixed equilibrium. This implies that in reinforcement learning with finite states, satisficing paths are guaranteed to exist regardless of whether agents select strategies based on current states or finite history records.

These results remain consistent with the findings in \cite{2023impossible, 2003uncoupled, 2012completely}, as no restriction is imposed on the update functions in this paper, unlike the regularity or uncoupled-ness conditions required in those studies. Conversely, these results provide theoretical support for stochastic update approaches \cite{2006stochastic-uncoupled, 2006regret-testing}. The existence of finite-length satisficing paths from any initial joint strategy to an equilibrium implies that: it is possible for stochastic algorithms to start from arbitrary initial joint strategy and to stop at some equilibrium, while keeping the player who has achieved best response not change its strategy.

\textbf{Open problems.} Theorem~\ref{theorem:existence} establishes a sufficient condition for the existence of grouped satisficing paths connecting any initial joint strategy to an equilibrium. This naturally leads to a question: What constitutes a necessary condition for Question~\ref{question}? More fundamentally, what is a complete necessary and sufficient condition to characterize the existence of such grouped satisficing paths?

Theorem~\ref{theorem:topology} and Theorem~\ref{theorem:existence} are established under the condition of $\epsilon=0$. This restriction arises because the application of the analytic extendability in Lemma~\ref{lemma:analytic} requires the use of certain equations to characterize best responses. So, does Theorem~\ref{theorem:existence} remain valid when $\epsilon>0$?

Grouped satisficing paths do not impose any neighborhood restriction on strategy selection for the player who has not achieved best response. This raises a question: does Theorem~\ref{theorem:existence} remain valid when non-best-responding players are restricted to select strategies only from a neighborhood of their current strategies? More fundamentally, what restriction can be imposed on non-best-responding players while still preserving the existence of grouped satisficing paths?

\textbf{Conclusion.} In brief, if the pure strategy game satisfies that $a)$ the strategy set is convex and compact, $b)$ the payoff function is continuous and partially analytic, $c)$ and any sub-game has an equilibrium, then for any initial joint strategy $s$, there exists a finite-length grouped satisficing path $(s^t)_{0 \leq t \leq T}$ where $s^0=s$ and $s^T$ is an equilibrium. In particular, any $N$-player game and any finite-state Markov game have the finite-length satisficing paths from any initial joint mixed strategy to some mixed equilibrium.

\bibliographystyle{acm}
\bibliography{reference}

\newpage
\appendix
\section*{Appendix} \label{section:appendix}

\section{Some proofs in Section~\ref{section:topologial-properties}} \label{appendix:topological-properties}
\begin{definition} \label{definition:dual-set}
In a pure strategy game $G$, for a joint strategy $s$, a subset of players $J \subset I$, and a player $i \in I \backslash J$, the dual $\epsilon$-best response set of player $i$ over $J$ is 
\begin{equation}
BRD_{\epsilon, i}(s, J) = \left\{x \in \prod_{j \in J} S_j \mid s_i \in BR_{\epsilon}(s_{-J \cup \{i\}}, x) \right\}.
\end{equation}
where $s_{-J \cup \{i\}} = (s_k)_{k \in I \backslash (J \cup \{i\})}$ denotes the strategy profile of all players outside $J \cup \{i\}$. In particular, the notation $BRD_{\epsilon, i}$ will be simplified to $BRD_i$ when $\epsilon=0$.
\end{definition}

The dual $\epsilon$-best response set $BRD_{\epsilon, i}(s, J)$ characterizes the strategic interdependence between player $i$ and the coalition $J$. If all players in $J$ adopt strategies $x \in BRD_{\epsilon, i}(s, J)$ while other players maintain their current strategies $s_{-J \cup \{i\}}$, then player $i$'s current strategy $s_i$ becomes an $\epsilon$-best response to the resulting strategy profile.

\begin{proposition} \label{proposition:dual-set-is-closed}
In a pure strategy game $G$, suppose that each $g_i$ is continuous. Then for a joint strategy $s$, a subset of players $J \subset I$, and a player $i \in I \backslash J$, the set $BRD_{\epsilon, i}(s,J)$ is closed.
\end{proposition}

\begin{proof}
By Definition~\ref{definition:best-response} and Definition~\ref{definition:dual-set}
$$
BRD_{\epsilon, i}(s, J) = \left\{x \in \prod_{j \in J} S_j \mid \forall t \in S_i, \,  g_i(s_i, s_{-J\cup\{i\}}, x) \geq g_i(t, s_{-J\cup\{i\}}, x) - \epsilon \right\}.
$$
Choose a limit point $x$ of $BRD_{\epsilon, i}(s,J)$, choose any $t \in S_i$, and construct a function
$$
h_t: \prod_{j \in J} S_j \to \mathbb{R}, \quad y \mapsto g_i(s_i, s_{-J\cup\{i\}}, y) - g_i(t, s_{-J\cup\{i\}}, y) + \epsilon.
$$
Since $g_i$ is continuous, $h_t$ is also continuous. Assume on the contrary that $h_t(x) < 0$, then $h_t^{-1}(-\infty, 0)$ is an open neighborhood of $x$. Since $x$ is a limit point of $BRD_{\epsilon, i}(s,J)$, there exists a point $y \in h_t^{-1}(-\infty, 0) \cap BRD_{\epsilon, i}(s,J)$. However
$$
g_i(s_i, s_{-J\cup\{i\}}, y) - g_i(t, s_{-J\cup\{i\}}, y) + \epsilon \geq 0.
$$
is a contradiction. As a result, $h_t(x) \geq 0$. Since $t$ is arbitrary
$$
\forall t \in S_i, \quad  g_i(s_i, s_{-J\cup\{i\}}, x) - g_i(t, s_{-J\cup\{i\}}, x) + \epsilon \geq 0.
$$
So $x \in BRD_{\epsilon, i}(s, J)$ which means $BRD_{\epsilon, i}(s, J)$ is a closed set.
\end{proof}

\begin{proposition} \label{proposition:graph-is-closed}
In a pure strategy game $G$, suppose that each $g_i$ is continuous. Then for any $j \in I$, the set
$$
A_j = \left\{s \in \prod_{i\in I} S_i \mid s_j \in BR_{\epsilon}(s_{-j})\right\}.
$$ is closed.
\end{proposition}

\begin{proof}
This proof is similar to the proof of Proposition~\ref{proposition:dual-set-is-closed}. Since $g_j$ is continuous, any limit point of $A_j$ belongs to $A_j$, which means $A_j$ is closed. 
\end{proof}

\subsection{Proof of Lemma~\ref{lemma:Baire}} \label{appendix:Baire}

{
\noindent\textbf{Lemma~\ref{lemma:Baire}.} \itshape In a pure strategy game $G$ with the group set $P$, suppose that each $S_i$ is compact and Hausdorff, and each $g_i$ is continuous. For a joint strategy $s$, if $N_{\epsilon}(s)$ is a local minimum, then

\begin{enumerate}
\item either for any joint strategy $t \in T_{\epsilon}(s)$ in the admissible subsequent joint strategy set, the group which has achieved $\epsilon$-best response in $s$ will still achieve $\epsilon$-best response in $t$, 
\item or there exists a non-empty open set $O \subset T_{\epsilon}(s)$ in the admissible subsequent joint strategy set satisfying:
\begin{enumerate}
\item For any joint strategy $t \in O$, there exists a group which has achieved $\epsilon$-best response in $s$ will not achieve $\epsilon$-best response in $t$.
\item There exists a group which has not achieved $\epsilon$-best response in $s$ will achieve $\epsilon$-best response in any joint strategy $t \in O$.
\end{enumerate}
\end{enumerate}
}

\begin{proof}
Without loss of generality, let $\vert P \vert = n+m$, group $p_1,\ldots, p_n$ achieve $\epsilon$-best response in the joint strategy $s$, while group $p_{n+1},\ldots,p_{n+m}$ not. Let
$$
J = \bigcup_{x=1}^n p_x, \quad K = \bigcup_{x=n+1}^{n+m} p_x.
$$
By assumption, all players in $J$ achieve $\epsilon$-best response in $s$. Construct a set
\begin{equation} \label{equation:define-A}
A = \bigcap_{j \in J} BRD_{\epsilon, j}(s, K).
\end{equation}
Since each $g_j$ is continuous, $BRD_{\epsilon, j}(s, K)$ is closed by Proposition~\ref{proposition:dual-set-is-closed}. $A$ is the intersection of finite closed sets, so $A$ is closed. 

Let
\begin{equation} \label{equation:define-V}
V = \prod_{k \in K} S_k.
\end{equation}
Since each $S_k$ is compact and Hausdorff, the product space $V$ is also compact and Hausdorff. According to Baire category theorem, $V$ is a Baire space. The set $V \backslash A$ is the complement of $A$ in $V$, so $V \backslash A$ is open. The open set in a Baire space is also a Baire space when it is non-empty, so the topological subspace $V \backslash A$ is a Baire space when it is non-empty.

Actually, any possible joint strategy following $s$ without violating the property of grouped $\epsilon$-satisficing path, has the same component of $s$ with index set $J$, and the entirety of their components with index set $K$ is $V$, namely
$$
T_{\epsilon}(s) = s_J \times V.
$$

As a result, $s_J \times A \subset s_J \times V$ is all admissible subsequent joint strategies which keep players in $J$ achieving $\epsilon$-best response, in other words, keep group $p_1, \ldots, p_n$ achieving $\epsilon$-best response. Since $N_{\epsilon} (s)$ is a local minimum, any joint strategy $t \in s_J \times (V \backslash A)$ must let at least one group $p \in  \{p_{n+1}, \ldots, p_{n+m}\}$ achieve $\epsilon$-best response. Otherwise $N_{\epsilon} (t) \leq N_{\epsilon} (s) - 1$, which is a contradiction with Definition~\ref{definition:local}.

Construct $m$ sets
\begin{equation} \label{equation:define-B}
B_{n + i} = \{ v \in V \mid w = (s_J, v), \, \forall x \in p_{n+i}, \, w_x \in BR_{\epsilon}(w_{-x}) \}, \quad i=1, \ldots, m.
\end{equation}
which means any joint strategy in $s_J \times B_{n+i}$ lets group $p_{n+i}$ achieve $\epsilon$-best response. Obviously
$$
B_{n+i} = \bigcap_{x \in p_{n+i}} \{ v \in V \mid w = (s_J, v), \, w_x \in BR_{\epsilon}(w_{-x}) \}, \quad i=1, \ldots, m.
$$
According to Proposition~\ref{proposition:graph-is-closed}, each set in the right-hand-side is closed, so $B_{n+i}$ is closed. So $B_{n+i} \cap (V \backslash A)$ is closed in the topological subspace $V \backslash A$, and
\begin{equation} \label{equation:cup-is-V-A}
\bigcup_{i=1}^m B_{n+i} \cap (V \backslash A) = V \backslash A.
\end{equation}

Two cases need to be discussed here.

\textbf{Case $1$.} $V \backslash A = \varnothing$. Then $A = V$ which means any possible joint strategy following $s$ without violating the property of grouped $\epsilon$-satisficing path, keeps group $p_1, \ldots, p_n$ achieve $\epsilon$-best response.

\textbf{Case $2$.} $V \backslash A \neq \varnothing$. According to Equation~\ref{equation:cup-is-V-A}, $B_{n+i} \cap (V \backslash A)$ is closed and $V \backslash A$ is a Baire space, there exists one $B_{n+i} \cap (V \backslash A)$ has a non-empty interior. So there exists a non-empty open set $C$ satisfying
$$
C \subset B_{n+i} \cap (V \backslash A).
$$
By Equation~\ref{equation:define-B}, any joint strategy in $s_J \times C$ lets group $p_{n+i}$ achieve $\epsilon$-best response. By Equation~\ref{equation:define-A}, Equation~\ref{equation:define-V}
and $C \subset V \backslash A$, any joint strategy in $s_J \times C$ makes at least one group $p \in \{p_1,\ldots, p_n\}$ no longer achieve $\epsilon$-best response. Obviously, $s_J \times C$ is an open set in $T_{\epsilon}(s)$.

\end{proof}

\begin{remark}
This lemma is very critical for Theorem~\ref{theorem:topology}. Usually, one point is defined as a closed set, so a closed set does not necessarily have a non-empty interior. In this lemma, Baire space and Baire theorem provide a guarantee to find a non-empty interior in finite closed sets when they satisfy some conditions.
\end{remark}

\subsection{Proof of Lemma~\ref{lemma:analytic}} \label{appendix:analytic}

{
\noindent\textbf{Lemma~\ref{lemma:analytic}.} \itshape In a pure strategy game $G$, suppose that each $S_i$ is a convex set in a topological vector space, and each $g_i$ is partially analytic. If there exists a non-empty open set $U \subset \prod_{i \in I} S_i$ satisfying that the player $k$ achieves best response in any joint strategy $s \in U$, then the player $k$ will achieve best response in any joint strategy $s \in \prod_{i \in I} S_i$.
}

\begin{proof}
Without loss of generality, let $I = \{1,\ldots,n\}$, and $k=1$. Since $U$ is open and non-empty, there exists a non-empty open set $O$
\begin{equation} \label{equation:define-O}
	O = \prod_{i \in I} O_i \subset U.
\end{equation}
where $O_i$ is open in $S_i$. Choose a joint strategy
$$
s = (s_i)_{i \in I}, \quad \forall i \in I, \quad s_i \in O_i.
$$
Then $s \in O$ which means the joint strategy $s$ makes player $k$ achieve best response. 

Induction needs to be used here.

\textbf{Head.} For any $t_k \in S_k$, construct a function
\begin{equation} \label{equation:define-function-f}
f: [0,1] \to \mathbb{R}, \quad x \mapsto g_k((1-x)s_k + x t_k, s_{-k}) - g_k(s_k, s_{-k}).
\end{equation}
Obviously, $f$ is an analytic function with respect to $x$ according to the assumption of $g_k$ and Definition~\ref{definition:analytic}. Consider
$$
t: [0,1] \to \prod_{i\in I} S_i, \quad x \mapsto ((1-x)s_k + x t_k, s_{-k}).
$$
Since each $S_i$ is a convex set in a topological vector space, the finite product space $\prod_{i \in I} S_i$ is a subspace in some topological vector space also. Because $t(0) = s \in O$, $O$ is open and $t(x)$ is a linear mapping, there exists an open neighborhood $(a,b) \subset \mathbb{R}$ of $0$ which satisfies
$$
\forall x \in [0,b), \quad t(x) \in O.
$$
According to Equation~\ref{equation:define-O} and the assumption of $U$
$$
\forall x \in [0, b), \quad(1-x)s_k + x t_k \in BR(s_{-k}).
$$
As a result
$$
\forall x \in [0, b), \quad g_k((1-x)s_k + x t_k, s_{-k}) = g_k(s_k, s_{-k}) \geq \sup_{r_k \in S_k} g_k(r_k, s_{-k}).
$$
So for any $x \in [0,b)$, $f(x) = 0$. Since $f$ is analytic, then
$$
\forall x \in [0,1], \quad f(x) = 0.
$$
So $t_k \in BR(s_{-k})$. Due to arbitrary $t_k$, $S_k = BR(s_{-k})$. Construct a set
$$
V_1 = S_k \times \prod_{i=2}^n O_i.
$$
So each joint strategy $s \in V$ makes player $k$ achieve best response.

\textbf{Recursion.} Assume 
$$
V_m = S_k \times \prod_{i=2}^m S_i \times \prod_{i=m+1}^n O_i.
$$
Each joint strategy $s \in V_m$ makes player $k$ achieve best response.

For any $t_k \in S_k$, $t_i \in S_i, i=2,\ldots, m+1$, construct a function
\begin{equation} \label{equation:define-function-h}
\begin{split}
h: & [0,1] \to \mathbb{R}, \\
& x \mapsto g_k(s_k, t_2, \ldots, t_m, (1-x)s_{m+1} + x t_{m+1}, s_{m+2}, \ldots, s_n) \\
& - g_k(t_k, t_2, \ldots, t_m, (1-x)s_{m+1} + x t_{m+1}, s_{m+2}, \ldots, s_n).
\end{split}
\end{equation}
Since each part of $h$ is analytic, $h$ is analytic. Since $O_{m+1}$ is open, there exists an open neighborhood $(c,d) \subset \mathbb{R}$ of $0$ which satisfies
$$
\forall x \in [0,d), \quad (1-x)s_{m+1} + x t_{m+1} \in O_{m+1}.
$$
So for any $x \in [0,d)$, $h(x) = 0$ according to the assumption of recursion. Due to analytic property, for any $x \in [0,1]$, $h(x) = 0$. So
$$
\forall t_k \in S_k, \quad g_k(t_k, t_2, \ldots, t_m, t_{m+1}, s_{m+2}, \ldots, s_n) = g_k(s_k, t_2, \ldots, t_m, t_{m+1}, s_{m+2}, \ldots, s_n).
$$
So
$$
S_k \subset BR(t_2, \ldots, t_m, t_{m+1}, s_{m+2}, \ldots, s_n).
$$
Construct a set
$$
V_{m+1} = S_k \times \prod_{i=2}^{m+1} S_i \times \prod_{i=m+2}^n O_i.
$$
Each joint strategy $s \in V_{m+1}$ makes player $k$ achieve best response.

By induction, each joint strategy $s \in \prod_{i\in I} S_i$ makes player $k$ achieve best response.

\end{proof}

\begin{remark}
The function $f$ in Equation~\ref{equation:define-function-f} is not similar to $h$ in Equation~\ref{equation:define-function-h}. $f$ is used to extend the best response property along $S_k$, while $h$ to extend the gap between different strategies in $S_k$ along $S_{m+1}$. The continuity of scalar multiplication in topological vector spaces and the extendability of analytic functions play a important role in the proof of this lemma.
\end{remark}

\section{Some proofs in Section~\ref{section:existence-of-paths}} \label{appendix:existence-of-paths}

{
\noindent\textbf{Proposition~\ref{proposition:finite-minimum}.} \itshape In a pure strategy game $G$ with the group set $P$, for any infinite-length path $(s^t)_{t \geq 0}$, there exists an index $v$ satisfying $N_{\epsilon}(s^v) = \inf_{t \geq 0} N_{\epsilon}(s^t)$. 
}

\begin{proof}
Construct $\vert P \vert +1$ sets
$$
A_i = \{t \mid N_{\epsilon}(s^t) = i\}, \quad i=0,\ldots, \vert P \vert.
$$
There exists a minimum $k$ making $\vert A_k \vert \neq \varnothing$. Choose an element $t \in A_k$. For finite joint strategies $s_0, s_1, \ldots, s_t$, there exists the first index $v$ making $N_{\epsilon} (s^v) = k$. So $N_{\epsilon}(s^v) = \inf_{i \geq 0} N_{\epsilon}(s^i)$. 
\end{proof}

\subsection{Proof of Corollary~\ref{corollary:continuous-game}} \label{appendix:continuous-game}

{
\noindent\textbf{Corollary~\ref{corollary:continuous-game}.} \itshape In a pure strategy game $G$, suppose that each $S_i$ is a convex compact set in a topological vector space and each $g_i$ is continuous and partially analytic. If for any $g_i$ and any $s_{-i} \in \prod_{j \neq i, j\in I} S_j$, the function $g_i(s_i, s_{-i})$ is quasi-convex with respect to $s_i$, then for any initial joint strategy $s$, there exists a finite-length satisficing path $(s^t)_{t=0}^T$ where $s^0=s$ and $s^T$ is an equilibrium.
}

\begin{proof}
There exists a theorem with respect to continuous games \cite{1952continuous-game}. In a pure strategy game $G=(I, (S_i)_{i\in I}, (g_i)_{i \in I})$, suppose that each $S_i$ is a convex compact set in a topological vector space, and each $g_i$ is continuous. If for each $g_i$ and any $s_{-i} \in \prod_{j \neq i, j\in I} S_j$, the function $g_i(s_i, s_{-i})$ is quasi-convex with respect to $s_i$, then $G$ admits at least one equilibrium.

As a result, for any subset $J \subset I$ and any joint strategy $s \in \prod_{i\in I} S_i$, construct a sub-game $\tilde{G} = (J, (S_i)_{i \in J}, (\tilde{g}_i)_{i \in J} )$ in which
$$
\tilde{g}_i: \prod_{i\in J} S_i \to \mathbb{R}, \quad w \mapsto g_i(s_{-J}, w).
$$
where $s_{-J} = (s_i)_{i\in I \backslash J}$ denotes the strategy profile of all players outside $J$. 

Function $\tilde{g}_i$ is also continuous, and for any $s_{-i} \in \prod_{j \neq i, j\in J} S_j$, the function $g_i(s_i, s_{-i})$ is quasi-convex with respect to $s_i$ as well. So $\tilde{G}$ has an equilibrium according to the theorem with respect to continuous games.

So all conditions of Theorem~\ref{theorem:existence} are satisfied. As a result, for any initial joint strategy $s$ in $G$, there exists a finite-length satisficing path $(s^t)_{t=0}^T$ where $s^0=s$ and $s^T$ is some equilibrium.
\end{proof}

\subsection{Proof of Corollary~\ref{corollary:simple-game}} \label{appendix:simple-game}

\begin{proposition} \label{proposition:product}
Suppose that real numbers $a_1, \ldots, a_n, b_1,\ldots, b_n$ belong to $[0,1]$, then
$$
\left\vert \prod_{i=1}^n a_i - \prod_{i=1}^n b_i \right\vert \leq \sum_{i=1}^n \vert a_i - b_i \vert.
$$
\end{proposition}

\begin{proof}
Obviously, the proposition is true when $n=1$. Assume the proposition is true when $n=k$, then
\begin{equation}
\nonumber
\begin{split}
\left\vert \prod_{i=1}^{k+1} a_i - \prod_{i=1}^{k+1} b_i \right\vert = & \left\vert \prod_{i=1}^{k+1} a_i - b_1 \prod_{i=2}^{k+1} a_i + b_1 \prod_{i=2}^{k+1} a_i  - \prod_{i=1}^{k+1} b_i \right\vert \\
\leq & \vert a_1-b_1 \vert \left\vert \prod_{i=2}^{k+1} a_i \right\vert  + \vert b_1 \vert \left\vert \prod_{i=2}^{k+1} a_i  - \prod_{i=2}^{k+1} b_i \right\vert \\
\leq & \vert a_1-b_1 \vert + \left\vert \prod_{i=2}^{k+1} a_i  - \prod_{i=2}^{k+1} b_i \right\vert.
\end{split}
\end{equation}
By assumption
$$
\left\vert \prod_{i=1}^{k+1} a_i - \prod_{i=1}^{k+1} b_i \right\vert \leq \vert a_1-b_1 \vert + \sum_{i=2}^{k+1} \vert a_i - b_i \vert.
$$
So the proposition is true when $n=k+1$.
\end{proof}

{
\noindent\textbf{Corollary~\ref{corollary:simple-game}.} \itshape In a mixed strategy game $G$, suppose that each $S_i$ is a finite set. Then for any initial joint mixed strategy $\sigma$, there exists a finite-length satisficing path $(\sigma^t)_{t=0}^T$ where $\sigma^0=\sigma$ and $\sigma^T$ is a mixed equilibrium.
}

\begin{proof}
Construct a pure strategy game $\tilde{G}=(I, (\Delta S_i)_{i\in I}, (\tilde{g}_i)_{i \in I})$ where $\Delta S_i$ is the set of all Borel probability measures over $S_i$, and
$$
\tilde{g}_i: \prod_{i \in I} \Delta S_i \to \mathbb{R}, \quad \prod_{i \in I} \sigma_i \mapsto \sum_{s \in \prod_{i \in I} S_i} g_i(s) \prod_{i\in I} \sigma_i(s_i).
$$
Since $S_i$ is a finite set, $\Delta S_i$ is actually a $(\vert S_i \vert -1)$-dimensional simplex in $\mathbb{R}^{\vert S_i \vert}$. So $\Delta S_i$ is a convex compact set in $\mathbb{R}^{\vert S_i \vert}$.

\textbf{Prove that $\tilde{g}_i$ is continuous.}

$g_i$ is bounded by
$$
M = \max_{s \in \prod_{j\in I} S_j} \vert g_i(s) \vert.
$$
The space $\prod_{i \in I} \Delta S_i$ can be endowed with a metric topology.
$$
\vert \sigma - \eta \vert = \sum_{i \in I} \vert \sigma_i - \eta_i \vert = \sum_{i \in I} \sum_{s \in S_i} \vert \sigma_i(s) - \eta_i(s) \vert, \quad \sigma,\eta \in \prod_{i \in I} \Delta S_i.
$$
Since the topology constructed by metric $\sum_{s \in S_i} \vert \sigma_i(s) - \eta_i(s) \vert$ is the same as the topology in $\mathbb{R}^{\vert S_i \vert}$, the definition above is reasonable.

For any $\epsilon > 0$, there exists $\delta = \epsilon / M$ such that for any $\vert \sigma - \eta \vert < \delta$
$$
\vert \tilde{g}_i(\sigma) - \tilde{g}_i(\eta) \vert \leq \sum_{s \in \prod_{i \in I} S_i} M \left\vert \prod_{i\in I} \sigma_i(s_i) - \prod_{i\in I} \eta_i(s_i) \right\vert. 
$$
According to proposition~\ref{proposition:product}
$$
\vert \tilde{g}_i(\sigma) - \tilde{g}_i(\eta) \vert \leq M \sum_{s \in \prod_{i \in I} S_i} \sum_{i \in I} \vert \sigma_i(s_i) - \eta_i(s_i) \vert = M \vert \sigma - \eta \vert < \epsilon. 
$$
So $\tilde{g}_i$ is continuous.

\textbf{Prove that $\tilde{g}_i$ is partially analytic.}

For any $j \in I$, $\sigma_j, \eta_j \in \Delta S_j$, $\sigma_{-j} \in \prod_{l \neq j, l\in I} \Delta S_l$, then
$$
\tilde{g}_i(x \sigma_j + (1-x) \eta_j, \sigma_{-j}) = \sum_{s \in \prod_{i \in I} S_i} g_i(s) (x \sigma_j(s_j) + (1-x) \eta_j(s_j)) \sigma_{-j}(s_{-j}).
$$
Obviously, $\tilde{g}_i(x \sigma_j + (1-x) \eta_j, \sigma_{-j})$ is a linear polynomial with respect to $x$, which means an analytic function with respect to $x$.

\textbf{Prove that any sub-game has an equilibrium.}

For any subset $J \subset I$ and any joint strategy $\sigma \in \prod_{i\in I} \Delta S_i$, construct a sub-game $\tilde{H} = (J, (\Delta S_i)_{i \in J}, (\tilde{h}_i)_{i \in J} )$ where
$$
\tilde{h}_i: \prod_{i\in J} \Delta S_i \to \mathbb{R}, \quad \eta \mapsto \tilde{g}_i(\sigma_{-J}, \eta).
$$
Construct functions
$$
h_i: \prod_{j \in J} S_j \to \mathbb{R}, \quad w \mapsto \sum_{v \in \prod_{j \in I \backslash J} S_j} g_i(w,v) \sigma_{-J}(v).
$$
So
$$
\tilde{h}_i(\eta) = \sum_{s \in \prod_{i \in J} S_i} h_i(s) \prod_{i\in J} \eta_i(s_i).
$$
which means that $\tilde{H}$ is the extended pure strategy game of a mixed strategy game $H = (J, (S_i)_{i \in J}, (h_i)_{i \in J} )$. 

There exists a theorem with respect to mixed strategy games \cite{1950bargain}. Each $N$-player game has a mixed equilibrium.

As a result, $H$ has a mixed equilibrium and $\tilde{H}$ has an equilibrium.

So all conditions of Theorem~\ref{theorem:existence} are satisfied. As a result, for any initial joint mixed strategy $\sigma$ in $G$, there exists a finite-length satisficing path $(\sigma^t)_{t=0}^T$ where $\sigma^0=\sigma$ and $\sigma^T$ is a mixed equilibrium.

\end{proof}

\subsection{Proof of Corollary~\ref{corollary:stationary-game}} \label{appendix:stationary-game}

\begin{proposition} \label{proposition:sum-binom}
For the real number $c \in (0,1)$ and the positive integer $p$, there exists
$$
\sum_{q \geq p, q \in \mathbb{N}} c^q \binom{q}{p} \leq  \frac{2 \sqrt{2\pi} p^{1/2}}{(-\ln c)^p} .
$$
when $p$ goes to infinity.
\end{proposition}

\begin{proof}
Since $\binom{q}{p} = q(q-1)\cdots (q-p+1) / p! \leq q^p / p!$
$$
\sum_{q \geq p, q \in \mathbb{N}} c^q \binom{q}{p} \leq \frac{1}{p!} \sum_{q \geq p, q \in \mathbb{N}} c^q q^p.
$$
The function $f(q) = c^q q^p$ has
$$
\frac{d}{dq} f = c^q q^{p-1} (q\ln c + p).
$$
Since $f(0^+) = f(+\infty) = 0$ and $f(q) \geq 0$, $f$ has only one maximal point $-p /\ln c$. $df/dq > 0$ when $0<q<-p/\ln c$, and $df/dq < 0$ when $-p/\ln c<q$. So
$$
\sum_{q \geq p, q \in \mathbb{N}} c^q q^p \leq \int_0^{\infty} f(q) dq + 2f\left(-\frac{p}{\ln c}\right).
$$
The form of the first term in the right-hand-side is similar to $\Gamma$ function.
$$
\Gamma(n+1) = \int_0^{\infty} t^n e^{-t} dt = n!.
$$
Actually
$$
\int_0^{\infty} f(q) dq = \left(-\frac{1}{\ln c}\right)^{p+1} \int_0^{\infty} e^{-t} t^p dq = \left(-\frac{1}{\ln c}\right)^{p+1} p!.
$$
So
$$
\sum_{q \geq p, q \in \mathbb{N}} c^q \binom{q}{p} \leq \left(-\frac{1}{\ln c}\right)^{p+1} + 2 \left(-\frac{1}{\ln c}\right)^{p} \frac{e^{-p} p^p}{p!} = \left(-\frac{1}{\ln c}\right)^{p} \left( -\frac{1}{\ln c} + 2\frac{e^{-p} p^p}{p!} \right).
$$
According to Stirling's formula $p! \sim \sqrt{2\pi p} p^p e^{-p}$, there exists $e^{-p}p^p/p! \sim \sqrt{2\pi p}$. So
$$
\sum_{q \geq p, q \in \mathbb{N}} c^q \binom{q}{p} \leq \left(-\frac{1}{\ln c}\right)^{p} 2\sqrt{2\pi} p^{1/2}, \quad p \to \infty.
$$

\end{proof}

{
\noindent\textbf{Corollary~\ref{corollary:stationary-game}.} \itshape In a stationary mixed strategy stochastic game $G$, suppose that each $S_i$ is a finite set. Then for any initial stationary joint mixed strategy $\sigma$, there exists a finite-length satisficing path $(\sigma^t)_{t=0}^T$ where $\sigma^0=\sigma$ and $\sigma^T$ is a mixed equilibrium.
}

\begin{proof}
According to the definition of $g_i$ and the finiteness of $X \times \prod_{i \in I} S_i$, $g_i$ is bounded by
$$
M = \max_{x \in X, s \in \prod_{i \in I} S_i} \vert g_i(x,s) \vert.
$$

Construct a pure strategy game $H=((I, X), (T_{i,x})_{(i,x) \in (I,X)}, (h_{i,x})_{(i,x) \in (I,X)})$. $(I,X)$ means a double index set, namely $(I,X) = \{(i,x) \mid i\in I, \, x \in X\}$. Since $\vert I \vert < \infty$ and $\vert X \vert < \infty$, $(I,X)$ is finite.
$$
T_{i,x} = \Delta S_i.
$$
Since $S_i$ is finite, $T_{i,x}$ is a simplex in some finite-dimensional Euclidean space. So $T_{i,x}$ is convex and compact.
$$
h_{i,x}: \prod_{(j,y) \in (I,X)} T_{j,y} \to \mathbb{R}, \quad \prod_{(j,y) \in (I,X)} \pi_j(y) \mapsto \mathbb{E}_{(\pi_j)_{j \in I}} \left[ \sum_{t \geq 0} \gamma_i^t g_i(x^t,s^t) \mid x_0=x \right].
$$
So
$$
h_{i,x}(\pi) = \mathbb{E}_{s \sim \pi} [g_i(x, s)] + \gamma_i \mathbb{E}_{s \sim \pi, y \sim P(x,s)} [h_{i,y}(\pi)].
$$
If $\pi \in \prod_{(j,y) \in (I,X)} T_{j,y}$ is considered as a constant, $h_{i,x}(\pi)$ will be viewed as a function with respect to $x$, denoted as
$$
h_{i}(\pi) : X \to \mathbb{R}, \quad x \mapsto h_{i,x}(\pi).
$$

The function space $\mathbb{R}^X$ is endowed with uniform metric topology, using
$$
\vert f \vert = \sup_{x\in X} \vert f(x) \vert, \quad f \in \mathbb{R}^X.
$$
as a norm in $\mathbb{R}^X$. Since $\mathbb{R}$ is complete, the uniform metric topology in $\mathbb{R}^X$ is complete.

Construct an operator
$$
op_{\pi}: \mathbb{R}^X \to \mathbb{R}^X, \quad f \mapsto \left\{(x, \mathbb{E}_{s \sim \pi} [g_i(x, s)] + \gamma_i \mathbb{E}_{s \sim \pi, y \sim P(x,s)} [f(y)]) \mid x \in X\right\}.
$$
For any $x \in X$, $p,q \in \mathbb{R}^X$, there exists
$$
\vert op_{\pi}(p)(x) - op_{\pi}(q)(x) \vert = \gamma_i \vert \mathbb{E}_{s \sim \pi, y \sim P(x,s)} [p(y) - q(y)] \vert \leq \gamma_i \sup_{y \in X} \vert p(y)-q(y) \vert = \gamma_i \vert p-q \vert.
$$
So
$$
\vert op_{\pi}(p) - op_{\pi}(q) \vert = \sup_{x\in X} \vert op_{\pi}(p)(x) - op_{\pi}(q)(x) \vert \leq \gamma_i \vert p-q \vert.
$$
Since $0 \leq \gamma_i < 1$, $op_{\pi}$ is a contraction mapping. 

According to contraction mapping theorem, there exists a unique fixed point. That is
\begin{equation} \label{equation:fix}
h_i(\pi) = op_{\pi}(h_i(\pi)).
\end{equation}
So $h_i(\pi)$ exists and is unique. As a result, $h_{i,x}$ is a well-defined function.

Obviously, $h_{i,x}(\pi)$ is bounded by
$$
\vert h_{i,x}(\pi) \vert \leq \sum_{t \geq 0} \gamma_i^t M \leq \frac{M}{1-\gamma_i}.
$$
This conclusion is always true no matter what $x$ or $\pi$ is. So for any $\pi \in \prod_{(j,y) \in (I,X)} T_{j,y}$
\begin{equation} \label{equation:bounded-h}
\vert h_i(\pi) \vert \leq \frac{M}{1-\gamma_i}.
\end{equation}

\textbf{Prove that $h_{i,x}$ is continuous.}

Since each $T_{i,x}$ is a metric topological space, the finite product space $\prod_{(i,x) \in (I,X)} T_{i,x}$ is also a metric topological space. The metric in this space is denoted as $\vert \pi - \sigma \vert$ where $\pi, \sigma \in \prod_{(i,x) \in (I,X)} T_{i,x}$.

For any $x \in X$, $p,q \in \mathbb{R}^X$, $\pi, \sigma \in \prod_{(i,x) \in (I,X)} T_{i,x}$, there exists
\begin{equation}
\nonumber
\begin{split}
\vert op_{\pi}(p)(x) - op_{\sigma}(q)(x) \vert \leq & \vert \mathbb{E}_{s \sim \pi} [g_i(x, s)] - \mathbb{E}_{s \sim \sigma} [g_i(x, s)] \vert \\
& + \gamma_i \vert \mathbb{E}_{s \sim \pi, y \sim P(x,s)} [p(y)] - \mathbb{E}_{s \sim \sigma, y \sim P(x,s)} [q(y)] \vert.
\end{split}
\end{equation}

Since
\begin{equation}
\nonumber
\begin{split}
\vert \mathbb{E}_{s \sim \pi} [g_i(x, s)] - \mathbb{E}_{s \sim \sigma} [g_i(x, s)] \vert = & \left\vert \sum_{s \in \prod_{j \in I} S_j} g_i(x, s) (\pi(x)(s) - \sigma(x)(s)) \right\vert \\
\leq & \sum_{s \in \prod_{j \in I} S_j} M \left\vert \prod_{k \in I} \pi_k(x)(s_k) - \prod_{k \in I} \sigma_k(x)(s_k) \right\vert \\
\leq & M \sum_{s \in \prod_{j \in I} S_j} \sum_{k \in I} \vert \pi_k(x)(s_k) - \sigma_k(x)(s_k) \vert \\
\leq & M C \vert \pi - \sigma \vert.
\end{split}
\end{equation}
where $C$ is an independent constant. The second $\leq$ inequality is guaranteed by Proposition~\ref{proposition:product}.

\begin{equation}
\nonumber
\begin{split}
& \vert \mathbb{E}_{s \sim \pi, y \sim P(x,s)} [p(y)] - \mathbb{E}_{s \sim \sigma, y \sim P(x,s)} [q(y)] \vert \\
= & \left\vert \sum_{s \in \prod_{j \in I} S_j, y \in X} (p(y) \pi(x)(s) P(x,s)(y) - q(y) \sigma(x)(s) P(x,s)(y))  \right\vert \\
\leq & \sum_{s \in \prod_{j \in I} S_j, y \in X} \vert \pi(x)(s) - \sigma(x)(s)  \vert \vert p(y) \vert P(x,s)(y) + \vert p(y)-q(y) \vert \sigma(x)(s) P(x,s)(y) \\
\leq & \sum_{y\in X} \left( \sum_{s \in \prod_{j \in I} S_j} \vert \pi(x)(s) - \sigma(x)(s)  \vert \vert p \vert \right) + \vert p-q \vert \sum_{s \in \prod_{j \in I} S_j, y \in X}  \sigma(x)(s) P(x,s)(y) \\
\leq &  C \vert \pi - \sigma \vert \vert p \vert \vert X \vert + \vert p-q \vert.
\end{split}
\end{equation}
So
\begin{equation} \label{equation:estimate-bound}
\vert op_{\pi}(p)(x) - op_{\sigma}(q)(x) \vert \leq (MC + \gamma_i C \vert p \vert \vert X \vert) \vert \pi-\sigma \vert + \gamma_i \vert p-q \vert.
\end{equation}

Here consider $\vert h_i(\pi) - h_i(\sigma) \vert$. According to Equation~\ref{equation:fix}, Equation~\ref{equation:bounded-h} and Equation~\ref{equation:estimate-bound}
\begin{equation}
\nonumber
\begin{split}
\vert h_i(\pi) - h_i(\sigma) \vert = & \vert op_{\pi}(h_i(\pi)) - op_{\sigma}(h_i(\sigma)) \vert \\
\leq & \left(MC + \gamma_i C \frac{M}{1-\gamma_i} \vert X \vert\right) \vert \pi-\sigma \vert + \gamma_i \vert h_i(\pi) - h_i(\sigma) \vert.
\end{split}
\end{equation}
So
$$
\vert h_i(\pi) - h_i(\sigma) \vert \leq \frac{MC}{1-\gamma_i} \left(1 + \frac{\gamma_i}{1-\gamma_i} \vert X \vert\right) \vert \pi-\sigma \vert.
$$
Since each element in the coefficient of the right-hand-side is independent of $\pi$ and $\sigma$, $h_i$ is continuous about $\pi$. As a result, $h_{i,x}$ is continuous about $\pi$.

\textbf{Prove that $h_{i,x}$ is partially analytic.}

For any $(j,y) \in (I, X)$, $\pi_j(y), \sigma_{j}(y) \in T_{j,y}$, $\eta \in \prod_{(k,z) \neq (j,y), (k,z) \in (I,X)} T_{k,z}$, let $\theta_a = (a \pi_j(y) + (1-a) \sigma_j(y), \eta)$, then
\begin{equation}
\nonumber
\begin{split}
h_{i,x}(\theta_a) = & \mathbb{E}_{\theta_a} \left[ \sum_{t \geq 0} \gamma_i^t g_i(x^t,s^t) \mid x_0=x \right] \\
= & \sum_{t \geq 0} \gamma_i^t \mathbb{E}_{x^0=x, \ldots, x^u \sim P(x^{u-1},s^{u-1}), s^u \sim \theta_a(x^u), \ldots, s^t \sim \theta_a(x_t) } g_i(x^t,s^t).
\end{split}
\end{equation}
Considering the boundary of the $k$-th term $c_k a^k$ in the polynomial with respect to $a$
$$
\vert c_k a^k \vert \leq M \sum_{t \geq k} \gamma_i^{t-1} \binom{t}{k} (2a)^k = RHS.
$$ 
Attention, the right-hand-side is the boundary of absolute sum of all possible coefficients of $a^k$ appeared in $h_{i,x}(\theta_a)$.

When $\gamma_i = 0$, $h_{i,x}$ is obviously analytic with respect to $a$.

When $0<\gamma_i<1$, according to Proposition~\ref{proposition:sum-binom}
$$
RHS \leq M(2a)^k \gamma_i^{-1} \frac{2\sqrt{2\pi} k^{1/2}}{(-\ln \gamma_i)^k} = \left( \frac{2a (2M)^{1/k} \gamma_i^{-1/k} (2\pi k)^{1/(2k)} }{-\ln \gamma_i} \right)^k , \quad k \to \infty.
$$
Obviously, $(2M)^{1/k} \gamma_i^{-1/k} (2\pi k)^{1/(2k)} \to 1$ when $k \to \infty$. So the radius of convergence is $-\ln \gamma_i / 2$, which is independent of $\pi_j(y)$ and $\sigma_j(y)$. This means that when $\vert a \vert < -\ln \gamma_i/2$, $h_{i,x}(\theta_a)$ is an analytic function with respect to $a$. Since the overall analyticity can be decomposed into the analyticity of segments, $h_{i,x}$ is analytic in the whole $[0,1]$.

\textbf{Prove that any sub-game has an equilibrium.}

Construct a group set
$$
A_i = \{(i,x) \mid \forall x \in X\} , \quad A = \{A_i \mid \forall i \in I\}.
$$
So $A_i$ is actually a real player in the original stochastic game who need choose a probability distribution over $S_i$ for each state in $X$. 

For any joint strategy $\pi \in \prod_{(i,x) \in (I,X)} T_{i,x}$, a subset $\Omega \subset A$, there exists $J \subset I$ satisfying
$$
\bigcup_{\omega \in \Omega} \omega = (J, X).
$$
Construct a sub-game $U = ((J,X), (T_{i,x})_{(i,x) \in (J,X)}, (u_{i,x})_{(i,x) \in (J,X)})$ where
$$
u_{i,x}: (T_{i,x})_{(i,x) \in (J,X)} \to \mathbb{R}, \quad \sigma \mapsto h_{i,x}(\pi_{- (J,X)}, \sigma).
$$
So
$$
u_{i,x}(\sigma) = \sum_{t \geq 0} \gamma_i^t \mathbb{E}_{x^0=x, \ldots, x^u \sim P(x^{u-1},s^{u-1}), s^u \sim (\pi_{(I \backslash J,X)},\sigma)(x^u), \ldots, s^t \sim (\pi_{(I \backslash J,X)},\sigma)(x_t) } g_i(x^t,s^t).
$$

Construct a stationary mixed strategy stochastic game $V=(J, (S_i)_{i \in J}, X, Q, (v_i)_{i\in I}, (\gamma_i)_{i\in I})$. 
$$
\forall s \in \prod_{i\in J} S_i, \quad \forall x \in X, \quad \forall y \in X, \quad Q(x,s)(y) = \sum_{w \in \prod_{i \in I \backslash J} S_i} \pi_{(I \backslash J,X)}(x) (w) P(x,s,w)(y).
$$
$$
\forall i \in J, \quad \forall s \in \prod_{i\in J} S_i, \quad \forall x \in X, \quad v_i(x,s) = \sum_{w \in \prod_{i \in I \backslash J} S_i} \pi_{(I \backslash J,X)}(x) (w) g_i(x,s,w).
$$

Induction needs to be used here.

\textbf{Head.} Choose $\sigma \in (T_{i,x})_{(i,x) \in (J,X)}$.

\begin{equation}
\nonumber
\begin{split}
\mathbb{E}_{x^0=x, s^0\sim \sigma(x^0)} v_i(x^0,s^0) = & \sum_{s^0 \in \prod_{i \in J} S_i} \sigma(x^0)(s^0) v_i(x^0,s^0) \\
= & \sum_{s^0 \in \prod_{i \in J} S_i} \sigma(x^0)(s^0) \sum_{w \in \prod_{i \in I \backslash J} S_i} \pi_{(I \backslash J,X)}(x^0) (w) g_i(x^0,s^0,w) \\
= & \mathbb{E}_{x^0=x, r^0\sim (\pi_{(I \backslash J,X)},\sigma)(x^0)} g_i(x^0,r^0)
\end{split}
\end{equation}
where $r^0 = (w, s^0)$.

\textbf{Recursion.} Assume for any $x^0 \in X$, there exists
$$
\mathbb{E}_{\ldots, x^u\sim Q(x^{u-1}, s^{u-1}), s^u\sim \sigma(x^u)} v_i(x^u,s^u) = \mathbb{E}_{\ldots, x^u\sim P(x^{u-1}, r^{u-1}), r^u\sim (\pi_{(I \backslash J,X)},\sigma)(x^u)} g_i(x^u,r^u).
$$
Consider the case $u+1$. To simplify notations, let
$$
\xi^u(x) = \mathbb{E}_{x^0=x, \ldots, x^u\sim Q(x^{u-1}, s^{u-1}), s^u\sim \sigma(x^u)} v_i(x^u,s^u).
$$
$$
\zeta^u(x) = \mathbb{E}_{x^0=x, \ldots, x^u\sim P(x^{u-1}, r^{u-1}), r^u\sim (\pi_{(I \backslash J,X)},\sigma)(x^u)} g_i(x^u,r^u).
$$
So
\begin{equation}
\nonumber
\begin{split}
\xi^{u+1}(x) = & \sum_{s^0 \in \prod_{i \in J} S_i} \sigma(x)(s^0) \sum_{x^{1} \in X} Q(x, s^0)(x^{1}) \xi^u(x^1) \\
= & \sum_{s^0 \in \prod_{i \in J} S_i} \sigma(x)(s^0) \sum_{x^{1} \in X} \xi^u(x^1) \sum_{w \in \prod_{i \in I \backslash J} S_i} \pi_{(I \backslash J,X)}(x) (w) P(x,s^0,w)(x^1) \\
= & \sum_{s^0 \in \prod_{i \in J} S_i} \sum_{w \in \prod_{i \in I \backslash J} S_i}  \sum_{x^{1} \in X} \sigma(x)(s^0) \pi_{(I \backslash J,X)}(x) (w) P(x,s^0,w)(x^1) \zeta^u(x^1)  \\
= & \zeta^{u+1}(x).
\end{split}
\end{equation}
So the case $u+1$ is valid no matter what $x^0=x$ is.

As a result
$$
u_{i,x}(\sigma) = \sum_{t \geq 0} \gamma_i^t \mathbb{E}_{x^0=x, \ldots, x^u \sim Q(x^{u-1},s^{u-1}), s^u \sim \sigma(x^u), \ldots, s^t \sim \sigma(x_t) } v_i(x^t,s^t).
$$
This means that the game $U$ is actually the extended form of the game $V$.

There exists a theorem with respect to stationary mixed strategy stochastic games \cite{1964equilibrium}. Each finite-state Markov game has a mixed equilibrium.

As a result, the game $V$ has a mixed equilibrium, which means that the game $U$ also has an equilibrium.

To sum up, all conditions of Theorem~\ref{theorem:existence} are satisfied. So in the game $H$, for any initial strategy $\pi$, there exists a finite-length grouped satisficing path $(\sigma^t)_{t=0}^T$ where $\sigma^0=\pi$ and $\sigma^T$ is an equilibrium.

Since the group set $A$ actually views a player in different state as a group, if a group achieves best response, the corresponding player will achieve best response in any state. Otherwise there exists a state where this player does not achieve best response.

As a result, for any initial stationary joint mixed strategy $\sigma$ in $G$, there exists a finite-length satisficing path $(\sigma^t)_{t=0}^T$ where $\sigma^0=\sigma$ and $\sigma^T$ is a mixed equilibrium.

\end{proof}

\subsection{Proof of Corollary~\ref{corollary:k-step-game}} \label{appendix:k-step-game}

{
\noindent\textbf{Corollary~\ref{corollary:k-step-game}.} \itshape In a $k$-step mixed strategy stochastic game $G$, suppose that each $S_i$ is a finite set. Then for any initial stationary joint mixed strategy $\sigma$, there exists a finite-length satisficing path $(\sigma^t)_{t=0}^T$ where $\sigma^0=\sigma$ and $\sigma^T$ is a mixed equilibrium.
}

\begin{proof}
Construct a stationary mixed strategy stochastic game $H$. 
$$
H=(I, (S_i)_{i \in I}, Y, Q, (h_i)_{i\in I}, (\gamma_i)_{i\in I}).
$$
$$
Y = X \times \left( \prod_{i\in I} S_i \right)^k.
$$
Obviously, $Y$ is finite, $\vert Y \vert < \infty$.
\begin{equation}
\nonumber
\begin{split}
Q: & Y \times \prod_{i\in I} S_i \to \Delta Y, \\
& \left(x, (s^{-t})_{t=1}^k\right) \times s \\
& \mapsto \left\{ \left( \left(y, (u^{-t})_{t=1}^k\right), P(x, s)(y) I[u^{-1}=s] \prod_{t=2}^k I[u^{-t}=s^{-t+1}] \right) \mid \left(y, (u^{-t})_{t=1}^k\right) \in Y \right\}. \\
\end{split}
\end{equation}
where $I[\cdot]$ equals $1$ if $[\cdot]$ is true else $0$.
$$
h_i: Y \times \prod_{i\in I} S_i \to \mathbb{R}, \quad \left(x, (s^{-t})_{t=1}^k\right) \times s \mapsto g_i(x, s).
$$
So the game $H$ is the extended form of the game $G$. When the runtime of the game $G$ exceeds $k$ time steps and the first $k$ steps of any path in $G$ are not be considered, each path in $G$ can be project onto a certain and unique path in $H$, and this projection is actually bijective.

According to Corollary~\ref{corollary:stationary-game}, for any initial stationary joint mixed strategy $\sigma$ in $H$, there exists a finite-length satisficing path $(\sigma^t)_{t=0}^T$ where $\sigma^0=\sigma$ and $\sigma^T$ is a mixed equilibrium in $H$. In turn, this conclusion is also true in $G$.

\end{proof}

\end{document}